\documentclass{article}
\usepackage{spconf,amsmath,graphicx,dcolumn,bm,color,qcircuit,braket,subcaption}
\usepackage{booktabs}       

\graphicspath{ {./images/} }
\usepackage[font=small,labelfont=bf,
  justification=justified]{caption}
\usepackage{hyperref}
\usepackage{setspace}

\title{Quantum Privacy Aggregation of Teacher Ensembles (qPATE) \\ for privacy preserving quantum machine learning}
\name{\begin{tabular}{@{}c@{}}
William Watkins$^{1*}$, Heehwan Wang$^{2*}$, Sangyoon Bae$^{2*}$,\\ Huan-Hsin Tseng$^4$, Jiook Cha$^2$, Samuel Yen-Chi Chen$^3$ \thanks{The first three authors$^*$ have an equal contribution. The views expressed in this article are those of the authors and do not represent the views of Wells Fargo. This article is for informational purposes only. Nothing contained in this article should be construed as investment advice. Wells Fargo makes no express or implied warranties and expressly disclaims all legal, tax, and accounting implications related to this article.}, Shinjae Yoo$^4$
\end{tabular}}
\address{Johns Hopkins University$^1$, Seoul National University$^2$, Wells Fargo$^3$, Brookhaven National Laboratory$^4$}
\begin{document}
%
\maketitle
\begin{abstract}
The utility of machine learning has rapidly expanded in the last two decades and presents an ethical challenge. 
Papernot et. al. developed a technique, known as Private Aggregation of Teacher Ensembles (PATE) to enable federated learning in which multiple \emph{teacher} models are trained on disjoint data sets.
This study is the first to apply PATE to an ensemble of quantum neural networks (QNN) to pave a new way of ensuring privacy in quantum machine learning (QML) models.
\end{abstract}
%
%
\section{Introduction}
\label{sec:intro}
\quad
Machine Learning (ML) is being utilized in a wide range of applications ~\cite{Sutskever2014SequenceNetworks, Wang_2021DeepFace, Szegedy2014GoingConvolutions, Voulodimos2018DeepReview, simonyan2014very, alipanahiDNADeepLearning} and often raises privacy and ethical concerns. 
%
%
%
For instance, privacy leaking is a major concern as demonstrated in large language models like GPT trained with sensitive texts~\cite{carlini2020extracting} and many applications deploy the model directly on the device, which would allow an adversary to use the model parameters directly to release private data. This is true especially in mobile applications in an effort to reduce communication with the server~\cite{deepLearningDP}. \emph{Differential Privacy} (DP) seeks to address these privacy concerns through its privacy-loss framework. One of the newer techniques to ensure differential privacy is known as \emph{Privacy-Aggregation of Teacher Ensembles}, i.e. PATE. It was pioneered by Nicholas Papernot et al. in 2017~\cite{papernot2017}. PATE provides strong privacy guarantees by training in a two-layer fashion. The training set is divided between $N$ \emph{teacher models}, who each train independently of each other. Then, the \emph{teacher ensembles} predict the labels of a disjoint training set through a noisy aggregation. The disjoint training set with newly aggregated noisy labels is used to train \emph{student}. The \emph{student}, never having access to the original training set or any \emph{teacher ensembles} model parameters, is deployed into the application. 

%

On the other hand, with the recent advancements of quantum computing hardware~\cite{bravyi2022future}, it is a natural next step to investigate \emph{Quantum Machine Learning} (QML).
%
%
%
Currently, there have been some investigations at the intersection of quantum computing and privacy-preserving machine learning~\cite{Song2021securinglearning, watkinsQML_DP}, yet to our best knowledge, there is no research implementing PATE with \emph{variational quantum circuits} (VQC).
This study aims to implement an ensemble of hybrid quantum-classical classifiers and train them using privacy-aggregation of teacher ensembles (PATE). After training, the student model will satisfy privacy loss limits.
Classical classifiers with PATE training will be used as controls in the study.
We demonstrate that the privacy-preserving VQC on MNIST handwritten digits has over 99\% accuracy with significant privacy guarantees.


\section{Privacy-Aggregation of Teacher Ensembles (PATE)}
\label{sec:PATE}

\subsection{\label{ssec:DP}Differential Privacy}
\quad
Differential Privacy (DP) has manifested as the standard tool for gauging privacy loss~\cite{dwork2006, dwork2011}. The notion of a privacy budget determines the amount of information that an adversary can extract. Information can be divided into two groups, \emph{general information} and \emph{private information}. The former refers to a general property of the dataset, whereas the latter refers to entry-specific information. DP puts limits on how much \emph{private information} can be ascertained from querying a database, or in the case of machine learning, a classifier~\cite{foundationDP}. 

DP is designated by two hyper-parameters, $\epsilon$ and $\delta$. The first parameter $\epsilon$ assigns the \emph{privacy budget} to the system and $\delta$ denotes the probability of leaking more information than allowed by the privacy budget~\cite{DPReview}.
%
The \emph{privacy budget} also known as the \emph{privacy loss}, $\mathcal{L}$, explicates the differences in the distributions characterized by two similar queries, $\mathcal{M}(x)$ and $\mathcal{M}(y)$. The privacy loss for a given observation, $\xi \in \mathrm{range}(\mathcal{M})$, quantifies the likelihood of observing $\xi$ from $\mathcal{M}(x)$ versus $\mathcal{M}(y)$.
$    \mathcal{L}^{(\xi)}_{\mathcal{M}(x) || \mathcal{M}(y)} = \ln\left(\frac{Pr[\mathcal{M}(x)=\xi]}{Pr[\mathcal{M}(y)=\xi]}\right)$.
%
Privacy-preserving algorithms are by definition non-deterministic, and an algorithm $\mathcal{M}$ is $(\epsilon, \delta)$-\emph{differentially private} if it satisfies,
    $Pr[\mathcal{M}(x) \in S] \leq \mathrm{exp}(\varepsilon) Pr[\mathcal{M}(y) \in S] + \delta$,
for all datasets $x,y$ as the domain of $\mathcal{M}$ with condition $\|x-y\|_1 \leq 1$ where $S$ denotes all possible events as outcomes of $\mathcal{M}$. 
Additionally, $\epsilon$ is the privacy loss for the randomized algorithm, and $\delta$ is the privacy cutoff, the probability that the model does not preserve privacy.
Differential privacy gives an upper-bound on the privacy loss and 
($\epsilon, \delta)$-differential privacy guarantees a privacy budget of $\epsilon$.

\subsection{\label{sec:DifferentialPrivacyMachineLearning}Differential Privacy in Machine Learning}

\quad For a machine learning context, the randomized algorithm, $\mathcal{M}$, can be cast as a training algorithm, which takes in a training dataset, $x \in X$, and leads a model, $\theta \in \Theta$~\cite{deepLearningDP}.

The most basic technique to ensure DP is the \emph{Gaussian Mechanism} as defined in \cite{foundationDP, mcmahanGeneralApproachDPTraining}, which is applied to machine learning in the following way. The gradient for all batches, $B$, is uniformly clipped to a length of $S$ and Gaussian noise is added to each element of the gradient, with a variance of $\sigma^2 S^2$, where $\sigma$ is the \emph{noise multiplier}. Higher levels of $\sigma$ produce less accurate yet more secure classifiers.
$\mathbf{g}_B \leftarrow \left[ \mathbf{g}_B*\mathrm{min}\left(1, \frac{S}{||\mathbf{g}_B||}\right) + \mathcal{N}(0,\sigma^2 S^2 I) \right]$.
The gradient contains all the information the classifier will ever see of the training data, so by manipulating that, one achieves privacy guarantees on the model. By clipping the batch loss gradient, one reduces the impact a given set of training datum can have on the training of the model parameters.
The hyperparamters associated with the \emph{Gaussian mechanism} are the noise multiplier, $\sigma$, and the $\ell_{2}$-norm cutoff, $S$ \cite{disparateImpact}.

More information about the Gaussian mechanism as applied to quantum machine learning can be found in \cite{watkinsQML_DP}. This modification to the optimizer algorithm can be applied to any ML optimizer (SGD, Adam, RMSprop, etc.).

Nicholas Papernot et al. employed a new technique to achieve differential privacy, in their models, called \emph{Privacy-aAggregation of Teacher Ensembles} (PATE).
Their method builds off of the work on knowledge aggregation and transfer~\cite{breiman1996, pathak2010, hamm2016}.
First an ensemble of models, known as the \emph{teachers}, are trained on disjoint private data. 
Then a \emph{student} model trains with public data, which is labeled by the output of the \emph{teacher ensemble}. The \emph{teacher ensemble} labels the data according to a noisy argmax voting system. 
PATE can provide privacy estimates that rival privacy-preserving techniques employed in real-life situations, such as Google's RAPPOR~\cite{erlingsson2014}.
Papernot et al. achieved an accuracy of 98\% for classifying MNIST data with $(2.04, 10^{-5})$-differential privacy, which was a tighter privacy bound than Abadi et al. MNIST classification~\cite{abadi2016}.

First, the training data $X$ and labels $Y$ are split into $n$ disjoint sets. Each teacher is trained on one of the disjoint subsets, $(X_i, Y_i)$ such that $(X, Y) = \bigsqcup_i^n \{(X_i, Y_i)\}$ to produce a model characterized by parameters $\phi_i$. The function $f_{\phi_i}(x)$ then predicts the label of the input data $x$. Next, the ensemble of teachers is queried about non-sensitive data, which aggregates the individual teacher predictions.

The aggregation of the \emph{teacher ensembles'} predictions forms the backbone of the PATE algorithm, by providing clear privacy guarantees.
Consider a classification task with $m$ different labels
$f(x) = \operatorname*{arg\,max}_{j \in [m]} \biggl\{ n_j(x) + Lap\biggl(\frac{1}{\gamma}\biggr) \biggr\}$
where $n_j(x) = |\{i:i\in[n],f_{\phi_i}(x)=j\}|$, or in other words, $n_j(x)$ is the number of teachers that predict label $j$ given an input $x$. $Lap(b)$ is a Laplacian distribution with location $0$ and scale $b$, and $\gamma$ is a privacy parameter~\cite{papernot2017}.
The addition of Laplacian noise introduces a privacy guarantee, by providing ambiguity in the number of teacher models that voted for a given label prediction.

Making predictions directly from $f$ would limit our privacy guarantee since our privacy loss would increase with each query. Additionally, `white-box' access breaks any claims of privacy on $f$. This reasoning provides a motivation to use the \emph{teacher ensemble} to train another model, the \emph{student}.

The \emph{student} model is released to the public, and it has a well-defined privacy loss because the student needs to only query the \emph{teacher ensemble} a finite number of times during training. Thus, the privacy loss does not increase as end-users query the \emph{student} model~\cite{papernot2017}.

Following Papernot et al., this study will use the momentum-accountant developed in Abadi et al. in 2016~\cite{abadi2016} to calculate the privacy loss of the \emph{student} model. One could calculate the privacy loss from each label query and then use Dwork's \emph{Composition Theorem} to calculate a total bound on the privacy loss, but this bound is too loose to be useful. Therefore, the momentum-accountant is used.

PATE has huge utility since there are no assumptions about the workings of the teachers or students. This is the first study to apply PATE to variational quantum classifiers.


\section{Quantum PATE (qPATE)}
\label{sec:QML}
\quad
We propose a hybrid quantum-classical framework that will train a \emph{student} classifier using a private-aggregation of \emph{teacher} classifiers.
A hybrid quantum-classical framework means that a quantum measurement is performed on a state generated by a quantum circuit. 
The quantum circuit output is compared to the ground truth on a classical computer, which will calculate the difference between the two using a \emph{loss function}.
The gradient of the loss function, with respect to the parameters of the quantum circuit, determines the direction that the parameters should evolve to reduce the loss and improve the accuracy of the machine. After updating the parameters in the circuit according to the loss gradient, the process repeats in an iterative fashion, \emph{training} the quantum circuit to replicate the desired function. Unlike Watkins et al.~\cite{watkinsQML_DP}, this study does not alter the gradients in the update-parameter step to achieve privacy guarantees. Instead, the labels in the training set of the student model are generated by the privacy-preserving assembly of teacher models. The `fuzziness' in the truth of the training labels creates an assurance of privacy.

\section{Variational Quantum Circuits}
\label{sec:VQC}
\quad
Modern deep learning relies on gradient-based optimization, while Quantum Machine Learning (QML) employs the parameter-shift rule for quantum model gradients ~\cite{schuld2019evaluating, mitarai2018quantum}. VQCs are fundamental for trainable quantum classifiers, with adaptable parameters updated during training~\cite{mitarai2018quantum, schuld2018circuit}. VQCs are resilient to noise, making them valuable for NISQ devices ~\cite{cerezo2020variational}, and have been used for computing chemical ground states~\cite{peruzzo2014variational}. The generic VQC consists of $R_y(\arctan(x_i))$ and $R_z(\arctan(x_i^2))$ rotations for \emph{variational encoding} and general single-qubit unitary gate $R(\alpha,\beta,\gamma)$ as the trainable components. CNOT gates are employed to establish qubit entanglement. The circuit's ultimate output is represented by the $\sigma_z$ measurement outcome

Our motivation arises from compelling evidence that hybrid VQC-deep neural networks (DNNs) outperform DNNs in terms of efficiency and convergence~\cite{sim2019expressibility,lanting2014entanglement,du2018expressive, abbas2020power}. Recent studies also show quantum circuits achieving faster learning and higher accuracy than classical counterparts, as in quantum CNN (QCNN) ~\cite{chen2022qcnn, chen2021qgcnn}, quantum long short-term memory (QLSTM) networks ~\cite{chen2022quantum} and natural language processing \cite{yang2020decentralizing}, highlighting their potential in advancing machine learning.

\section{Experiment Methodology}
\label{sec:method}
\quad
This study demonstrates a privacy-preserving hybrid classical-quantum classifier using the PATE technique developed by Papernot et al.~\cite{papernot2017}. We illustrate the efficacy of PATE with hybrid VQC-DNNs (quantum PATE) by comparing it to PATE with DNNs (classical PATE). 
A binary classification of MNIST digits is used as a benchmark for quantum PATE, which was also used in Watkins et al.~\cite{watkinsQML_DP} in their study of privacy-preserving classifications with quantum circuits.

\subsection{MNIST Binary Classification}

\quad
MNIST is a representative benchmark dataset in deep learning and quantum machine learning studies ~\cite{deng2012mnist}. MNIST includes $28 \times 28$ pixel black-white images of handwritten digits. Because of the computational complexity of simulating large quantum systems, the MNIST classification task is reduced to a binary classification of distinguishing the handwritten digits of `0' and `1.' The original inputs are padded with additional zeros to be images of size $32 \times 32$.

\subsection{Model Architecture and Hyperparameters}
\begin{figure}[htbp]
\begin{center}
\begin{minipage}{10cm}
\includegraphics[width=0.8\textwidth]{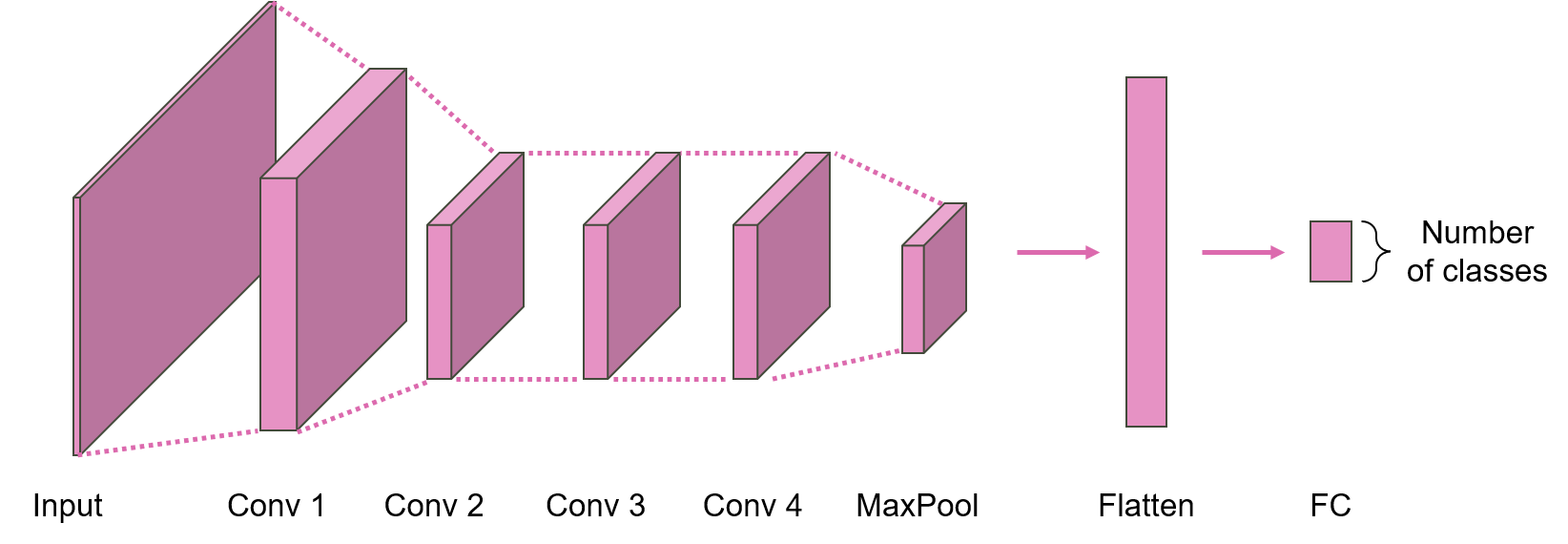}
\end{minipage}
\end{center}
\caption[Classical PATE network architecture]{{\bfseries Classical PATE network architecture.}
 Classical PATE uses four convolution blocks.
 }
\label{Fig:cPATE_architecture}
\end{figure}

\begin{figure}[htbp]
\begin{center}
\begin{minipage}{10cm}
\includegraphics[width=0.8\textwidth]{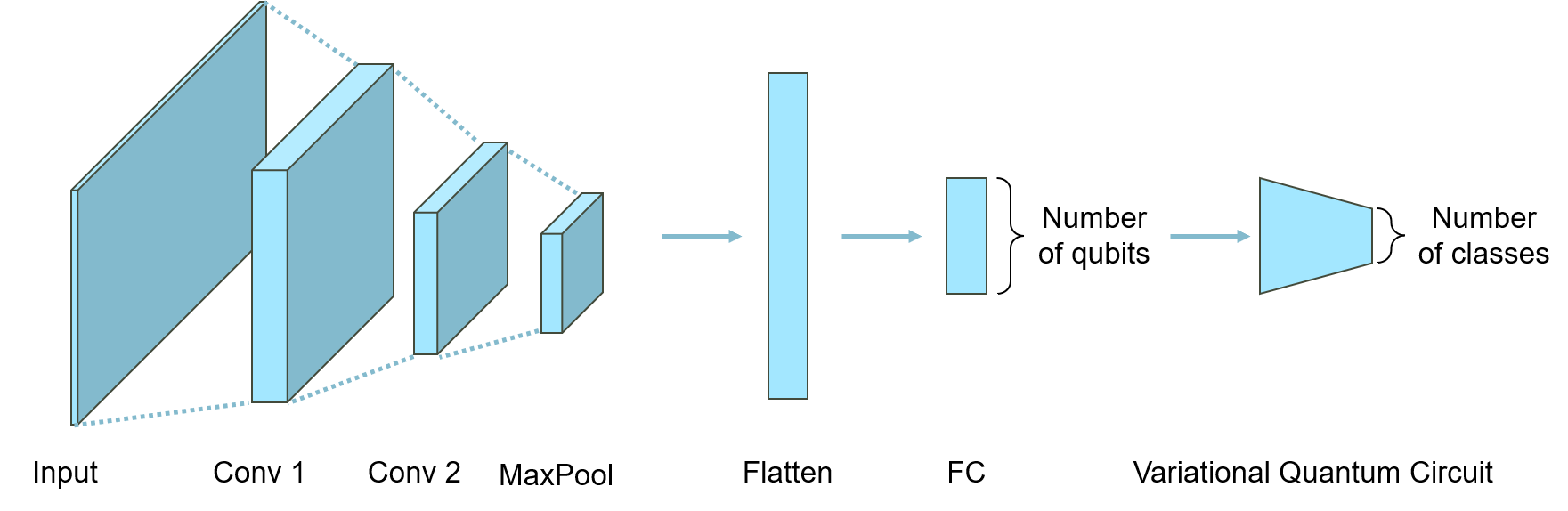}
\end{minipage}
\end{center}
\caption[Quantum PATE network architecture]{{\bfseries Quantum PATE network architecture.}
 Quantum PATE uses two convolution blocks with the additional VQC block.}
\label{Fig:qPATE_architecture}
\end{figure}

\quad
As part of the investigation into differentially private QML, classical PATE and quantum PATE are compared. In classical PATE, both teacher and student networks are made out of four convolution blocks. As depicted in Figure~ \ref{Fig:cPATE_architecture}, the former two convolution blocks consist of a 3x3 convolutional layer, batch normalization, and ReLU activation, and the latter two convolution blocks consist of a 1x1 convolutional layer, batch normalization, and ReLU activation. On the other hand, in quantum PATE (Figure \ref{Fig:qPATE_architecture}), both teacher and student networks are made out of two convolution blocks and an additional VQC block. Two convolution blocks consist of a 3x3 convolutional layer, batch normalization, and leaky ReLU activation.

\begin{figure}[htbp]
\begin{center}
\begin{minipage}{10cm}
\includegraphics[width=0.8\textwidth]{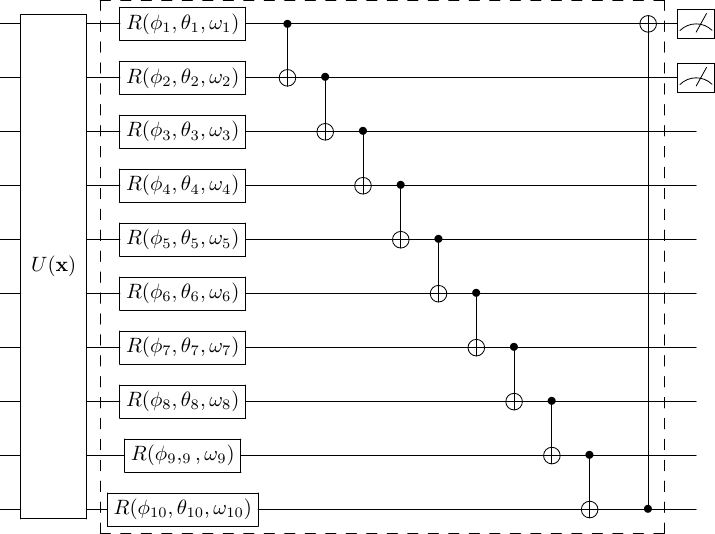}
\end{minipage}
\end{center}
\caption[VQC block for MNIST classification]{{\bfseries VQC block for MNIST classification.}
 The VQC block encodes latent embeddings from convolution blocks within quantum PATE into quantum states represented by $10$-qubits. $U(\mathbf{x})$ denotes the quantum algorithm for angle encoding. $\phi_{i}$, $\theta_{i}$, and $\omega_{i}$ are the parameters to optimize. The dashed box denotes one subcircuit of the VQC block that is repeated two times. The dial to the far right represents that the circuit has two outputs. The expectation of $\sigma_z$ is measured on two qubits.
 }
\label{Fig:MNIST_VQC1}
\end{figure}

The VQC block within quantum PATE has two subcircuits for \textit{variational encoding}, following two subcircuits for \textit{angle encoding} ~\cite{mottonen2005transformation, Schuld2018InformationEncoding}. After $512$-dimensional latent embeddings from classical convolution blocks are reduced to $10$-dimensional latent embeddings, two angle encoding subcircuits encode these latent embeddings into a $10$-qubit state. Subsequently, the outputs from angle encoding pass through two variational encoding subcircuits, each consisting of a rotation gate and a CNOT (controlled-NOT) gate as described in Figure \ref{Fig:MNIST_VQC1}. Since the rotation gate assigned to each qubit is parameterized with three parameters, and the subcircuit consists of 10 qubits, there are $3 \times 10 \times 2 = 60$ parameters in the whole circuit.

%

The experiments are characterized by the hyperparameters of the neural network training process: the optimizer, number of epochs, number of training samples, learning rate, batch size,  and weight penalty. When differentially private optimizers are used, the additional hyperparameters needed are the noise multiplier and $\epsilon$. After preliminary experiments, the AdamW optimizer was selected for use in all of the experiments presented in this paper. Most of the model's hyperparameters were the same for both classical PATE and quantum PATE; the learning rate was set to $0.001$, while the number of train, validation, and test datasets were 1,000, 100, and 100, randomly assigned to each teacher and student networks respectively. The batch size was $64$ with a weight decay $10^{-4}$. 

An $\epsilon$ was calculated from the DP hyperparameters, $S, \sigma$. Because all tasks were classifications, cross-entropy was used as the loss function for all training. According to ~\cite{DPReview, foundationDP}, the probability of breaking $\epsilon$-DP should be $\delta \sim \mathcal{O}(1/n)$ for $n$ samples. A $\delta$ larger than $1/n$ always will be able to satisfy DP simply by releasing $n\delta$ complete records. Therefore, $\epsilon$ was determined by hyperparameter choice, and $\delta$ was set to be $10^{-5}$ for the entire study. Various noise multipliers were employed for comparing differentially private networks, with the ``varies'' noise parameter indicating the use of multiple noise values in both classical PATE and quantum PATE experiments. $\epsilon$ also varied across DP experiments as it directly depends on the noise multiplier.

%

%
%


\section{Results}
\label{sec:results}

\begin{table}[htb]
\label{VBD_table_pesq}
\centering 
\begin{tabular}{cccc}
\toprule
$\epsilon$ & $\delta$ &classical PATE & quantum PATE\\
\cmidrule(r){1-4}
 0.01 & $10^{-5}$ & 0.534 $\pm$ 0.0992 & \textbf{0.688} $\pm$ 0.0163\\ 
 0.1 & $10^{-5}$ &  0.985 $\pm$ 0.0215 &  \textbf{0.992} $\pm$ 0.0098 \\
 1.0 & $10^{-5}$ &  \textbf{0.997} $\pm$ 0.0046 &  0.99 $\pm$ 0.0134 \\
 10.0 & $10^{-5}$ &  \textbf{0.997} $\pm$ 0.0046 &  0.991 $\pm$ 0.0137 \\
\bottomrule
\end{tabular}
\caption{Results from binary MNIST classification. Accuracies of classical PATE and quantum PATE after 20 epochs. The private quantum classifier is more accurate and successful for \bm{$\epsilon$} between 0.01 and 0.1. The number of teachers is set as 4.}
\label{table:eps}
\end{table}




\begin{figure}[htbp]
\begin{center}
\begin{minipage}{10cm}
\includegraphics[width=0.9\textwidth]{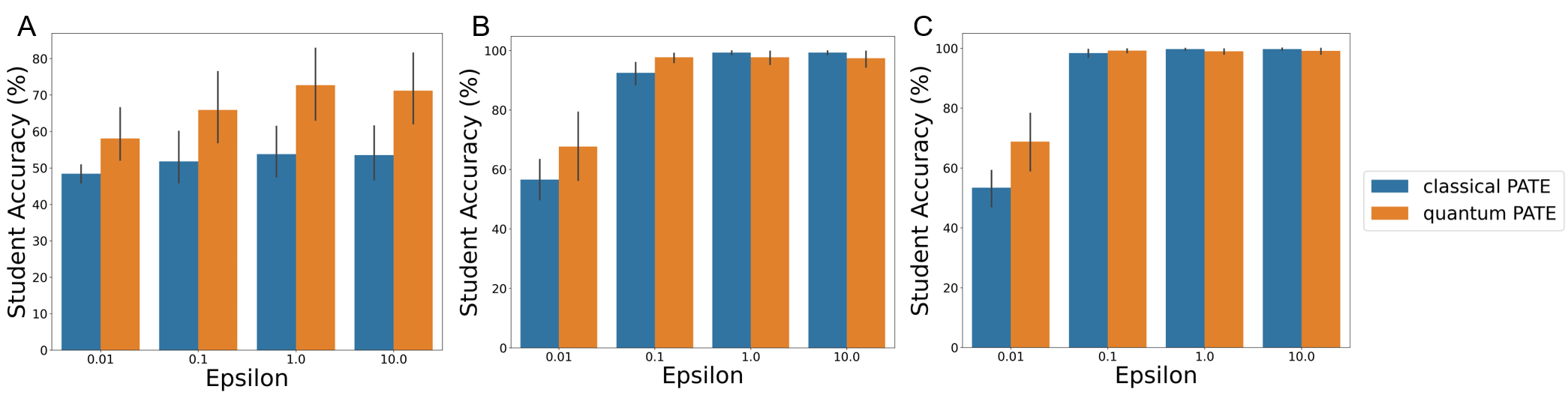}
\end{minipage}
\end{center}
\vskip -0.2in
\caption[acc-epsilon plot]{{\bfseries Accuracy vs. Epsilon Plot for 4 Teachers in classical PATE and quantum PATE.}
We averaged the results of 10 experiments, and the error bar denotes the standard deviation. (A) is the result of 1 epoch training, (B) is the result of 10 epoch training, (C) is the result of 20 epoch training.}
\label{Fig:acc_epsilon}
\end{figure}

\quad
Since $\epsilon$ measures the amount of privacy loss at a differential change in data, the lower the $\epsilon$ value, the stronger the privacy protection is. As seen in Table~\ref{table:eps}, quantum PATE showed 28.84\% higher performance than classical PATE at $\epsilon$ value of 0.01. While at $\epsilon$ value of greater than 0.1, the prediction performance of both classical PATE and quantum PATE was comparable. 

Furthermore, we observed that quantum PATE achieved quite a good test performance only after 1 epoch of training at various $\epsilon$ values (Figure \ref{Fig:acc_epsilon}). After longer epochs of training, the accuracy of classical PATE and quantum PATE became comparable at high $\epsilon$ values, i.e., 0.1, 1, 10. However, at a low $\epsilon$ value, quantum PATE still outperforms classical PATE. These results indicate that quantum PATE outperformed classical PATE in terms of accuracy and convergence while preserving high privacy protection compared to classical PATE.

\section{Conclusion}
\label{sec:conclusion}
\quad
This is the first implementation of a hybrid quantum-classical classifier that preserves privacy through privacy-aggregation of teacher ensembles (PATE). Furthermore, our results demonstrate a quantum advantage since the models are of the same order of complexity. 

On the other hand, we highlight the potential of a hybrid quantum-classical framework in implementing accurate and privacy-protecting machine learning algorithms. Though it's crucial to attain high prediction accuracy while keeping $\epsilon$ values low to ensure privacy protection, the trade-off between prediction accuracy and $\epsilon$ value makes it hard to achieve these goals simultaneously. Nevertheless, our results conclusively demonstrate that a hybrid approach, combining Variational Quantum Circuits (VQC) with Deep Neural Networks (DNNs), has the capability to significantly enhance prediction accuracy at low $\epsilon$ values when compared to using only DNNs. In future works, we will test that our results can be generalized to more complex image classification tasks, e.g., CIFAR10, ImageNet21k.

\vfill\pagebreak


\clearpage
\begin{spacing}{0.6}
\footnotesize
\bibliographystyle{IEEEbib}
\bibliography{main}
\end{spacing}

\end{document}